\documentclass[twocolumn,showpacs,preprintnumbers,amsmath,amssymb]{revtex4}

\usepackage{graphicx}
\usepackage{dcolumn}
\usepackage{bm}

\begin{document}


\title{A new representation of the bulk current in the quantum Hall effect regime}

\author{Josef Oswald}

\affiliation{%
Institute of Physics, University of Leoben, Franz Josef Str.18, A-8700 Leoben, Austria, e-mail: josef.oswald@unileoben.ac.at}%

\date{\today}

\begin{abstract}
In preceding papers a Landauer-B\"uttiker type representation of
bulk current transport has been successfully used for the
numerical simulation of the magneto transport of 2-dimensional
electron systems in the high magnetic field regime. In this paper
it is demonstrated, that this representation is in full agreement
with a treatment of the bulk current transport as a tunneling
process between magnetic bound states. Additionally we find a
correspondence between our network representation and the bulk
current picture in terms of mixed phases mapped on a checkerboard:
At half filled Landau level (LL) coupled droplets of a quantum
Hall (QH) liquid phase and coupled droplets of an insulator phase
phase exist at the same time, with each of them occupying half of
the sample area. Removing a single electron from to such a QH
liquid droplet at half filling completes the QH transition to the
next higher QH plateau. Adding a single electron to such a droplet
at half filling completes the QH transition to the previous lower
QH plateau. As a consequence, the sharpness of the QH plateau
transitions on the magnetic field axis depends on the typical size
of the droplets, which can be understood as a measure of the
disorder in the sample.
\end{abstract}

\maketitle

\section{\label{sec:level1}Introduction}

Even more than 20 years after the discovery of the integer quantum
Hall effect there are still controversial discussions about the
origin of this phenomenon. In particular the transition regime
between plateaus gained increasing attention during the last two
decades. While it is commonly accepted, that the plateau values of
the integer quantum Hall effect (IQHE) can be explained by the
edge channel (EC) picture\cite{Buettiker1,Chklovskii,Hirai,
Christen}, the transition regime between plateaus is believed to
be driven by a different mechanism in the bulk of the sample. A
well defined scaling behavior of the temperature dependence is
seen as an evidence for a quantum phase transition in the vicinity
of the QHE plateau transitions \cite{Wei1, Koch1, Koch2, Wei2,
Hohls} (for a review see also Refs. \cite{1,2}). While the
existence of a formal equivalence of the edge channel picture and
the bulk current picture is proposed by Ruzin et al.\cite{9}, the
investigations of non-equilibrium situations like selective edge
channel population, AC-characteristics and non-linear transport
give evidence, that edge channels should be more than just a
different representation of bulk currents. Since more than two
decades the fundamental question concerning edge vs. bulk in the
QHE gets permanent attention even until now, as can be seen from a
representative selection of references: Starting with the
discovery of the IQHE \cite{Klitzing} and first systematic
experiments investigating different contact configurations
\cite{Stiles1, Stiles2}, continuing with the introduction of edge
channels by B\"uttiker \cite{Buettiker1} and the enormous number
of examples for dealing with the edge vs. bulk problem and making
use of the EC picture for modeling and interpretation of
experiments \cite{Kane,Streda,13,Jain1,11,12,Komiyama6, vanWees2,
vanWees3, 14, Alphenaar, Martin, Muller2, Geim1, Simmons1, Ford,
Son, Komiyama4, Komiyama5, Alphenaar2, Komiyama7, Keilmann,
Maslov, Lee, Mueller, Geim2, Komiyama9, Peck, Murase1, Kotthaus1,
Kristensen, vanWees1, vanHaren1, Zozulenko, Johnson, Komiyama2,
Yahel, Rahman, Mani2, Gagel, oswald3, oswald4, Kao, Komiyama1, 17,
Bird1, Shahar2, Tsemekhman, Mani1, oswald5, oswald6, 5, 4,
Komiyama1a, Furusaki, 3, Komiyama1b, oswald7, oswald8, oswald9,
oswald10, 6, probing5, Komiyama8, oswald2, probing2, probing3,
probing4, Ahlswede1,Komiyama3a, network, Machida2, Ahlswede2,
probing1,8, oswald11}, still a remarkable number of most recent
papers on this topic can be found \cite{Ensslin1, Komiyama2a,
Komiyama3, Blaauboer, Takashina, Buttiker_hanbury1, Peled1,
Machida1, Komiyama_distr, Arai, Cresti, Buttiker_asymm,
Buttiker_hanbury2, Peled2, Hehl, Deviatov, 16, Seo, Ensslin2}.
Concerning experimental probing of edge channels, major
improvements of the experimental techniques have been achieved in
the last few years \cite{Keilmann, Murase1, vanHaren1, Yahel,
probing5, probing2, probing3, probing4,Ahlswede1, Ahlswede2,
probing1, Ensslin2}. A very fundamental aspect of edge channels is
the drastic enhancement of the phase coherence length of electrons
within edge states. Major work on this topic has been done by
Komiyama et al \cite{Komiyama2a}.

Concerning modeling of quantum transport, one of the very first
attempts can be attributed to R.Landauer\cite{Landauer1,
Landauer2}. A major step forward was achieved by M. B\"uttiker by
introducing the so-called Landauer-B\"uttiker (LB) formalism and
the EC-picture of the QHE\cite{Buettiker1}. For modeling quantum
transport and localzation in the bulk, mainly network models on
the basis of the Chalker Coddington (CC) network \cite{CD} are
used. However, a model, which generates data in terms of voltages
and resistances, which would allow a direct comparison which
experimental data for realistically shaped samples had not been
developed so far. In order to do so, an approach to combine EC
transport and bulk transport has been made some time ago
\cite{3,4}. Subsequently that model has been expanded to a
network\cite{network}, which is also the subject of this paper.
Although the basic idea of our network is common with the basic
idea of the CC network, our handling of the nodes is substantially
different: In contrast to a CC network our network does not use a
transfer matrix for amplitudes and phases. We use transmission by
tunneling, but incorporating the effect of tunneling according to
the LB formalism. The nodes are described by a back scattering
function $P$, which is the ration of reflection and transmission
coefficient $R/T$.

\begin{figure}
\includegraphics{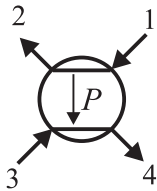}
\caption{\label{node1} Node of the network with two incoming and
two outgoing channels. The channels $1 \rightarrow 2$ and $3
\rightarrow 4$ are treated like edge channels with back
scattering, where $P=R/T$ according to the Landauer- B\"uttiker
formalism.}
\end{figure}

In Fig.\ref{node1} a single node of the network is shown, which
transmits potentials from the incoming to the outgoing channels.
The transmitted potentials of the outgoing channels as a function
of $P$ is calculated as follows \cite{network}:

\begin{subequations}
\label{eq0}
\begin{equation}
\mu_2 = (\mu_1 + P \cdot \mu_3)/(1+P)
\end{equation}
\begin{equation}
\mu_4 = (\mu_4 + P \cdot \mu_1)/(1+P)
\end{equation}
\end{subequations}

\begin{figure}
\includegraphics{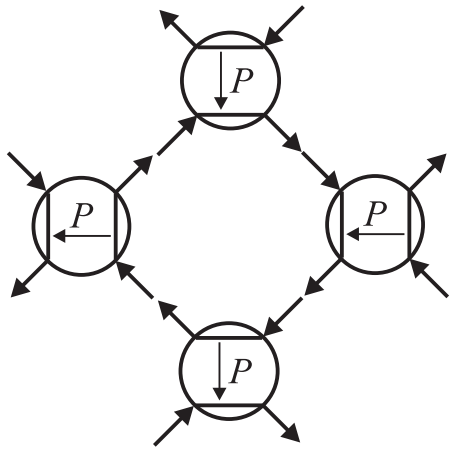}
\caption{\label{node2} Arrangement of the nodes for building the
minimal physical element of a network, which is the closed loop of
a so called magnetic bound state (see below) }
\end{figure}

In Fig.\ref{node2} it is shown, how a network is build by
arranging the nodes. The minimum network forms a loop, which
corresponds physically to a single magnetic bound state. A
periodic continuation of this network leads to further adjacent
loops, which get coupled by the nodes. In this way the network
transmits potentials and the potential distribution is calculated
by an appropriate iteration procedure. Currents are obtained only
after obtaining the solution for the potential distribution within
the whole network. Because of using just reflection and
transmission coefficients so far, phase coherence between adjacent
loops is not explicitly included. Despite this fact, this network
allows to simulate IQHE experiments with realistic sample geometry
in almost all details\cite{network, 8}. The key-point of this
approach is the representation of the back scattering function
$P$, which will be considered in more detail in the main part of
this paper.

Considering the nodes of the network, they can be understood as
´elementary´ QH - samples with a single EC pair experiencing back
scattering, which is described by the function $P$. Originally
this function had been used as representative for a QHE sample as
a whole and it had to be defined for each LL separately as
$P=P(\Delta\nu)$\cite{3,4}. $\Delta \nu$ is the filling factor of
the corresponding LL relative to half filling. The total sample
behavior was obtained by summing up the components of the
conductance tensor of all LLs. In \cite{3} it had been shown, that
in order to meet all symmetry relations found in experimental data
\cite{5, Shahar2}, $P( \Delta \nu )$ has to be an exponential
function $P( \Delta \nu )= exp (-\Delta \nu / k)$. On this
background this function has to be seen as a semi-empirical
function so far. Nevertheless, just assuming the existence of this
function allowed to explain already quite a number of universal
features of the IQHE without referring to a special choice of the
pre-factor $k$ in the exponent. The function $P( \Delta \nu )$
also allows to obtain a scaling behavior by considering the factor
$k$ as a temperature dependent function $k(T)$. In \cite{3} it has
been shown, that any not necessarily known function $k(T)$ is
mapped out directly by the temperature scaling function for
$R_{xx}$ and $R_{xy}$.

The basic idea, which subsequently led to the usage of $P(\Delta
\nu )$ also within a network approach, is that current transport
at high magnetic fields in the bulk region may happen via
tunneling between magnetic bound states. Such bound states are
considered to be physically equivalent to ECs. Thus, these bound
states are closed loops of directed channels and are created in
real samples by smooth potential fluctuations at high magnetic
fields. Tunneling between such loops happens preferably near the
saddle points of the random potential. The most important
ingredient of our network is that the tunneling current can be
handled as a back scattering process in the EC-picture like
already outlined above. This makes our network approach essential
different from the Chalker-Coddington model\cite{CD} and all
network approaches derived from this\cite{Cho, hirarchnetwork1,
hirarchnetwork2}. Just to clarify, the CC network maintains
coherence on the entire network, while the largest coherent
element in our network is a single loop, which at the same time
defines the size of a network period $L$. If we consider the
in-elastic scattering length $L_{in}$ of the electron system and
the size of the sample $L_p$, then the CC network deals with the
regime $L<L_p\leq L_{in}$, while our network considers the regime
$L_{in}\leq L < L_p$. From this point of view our network does not
necessarily contradict the CC-network, but addresses a different
regime. As will be shown in this paper, the tunneling process is
well represented by the back scattering function $P(\Delta\nu)$
used in \cite{network}. It can be calculated locally at the
designated grid points from the magnetic field and the local
carrier density. On this basis it is possible to model the shape
the sample by shaping the lateral carrier distribution. By doing
so, it has been shown for the first time, that the influence of
sample geometry (such as contact arms) and carrier density
inhomogeneities (like introduced by gate electrodes) can be
successfully addressed on the basis of a network model in full
agreement with experimental results \cite{network,8}. While the
main task of\cite{network} was to present the technical aspects
and examples for applications of the network model, the main
purpose of this paper is to provide the theoretical and physical
back ground. Since the network model is based on the idea of
tunneling between magnetic bound states at saddles of the interior
potential landscape, a systematic study of this tunneling process
will be presented. Also the interesting question, whether or not
there is a correspondence between the edge channel picture and the
bulk current picture in terms of mixed phases, will also be
addressed.

\section{\label{sec:level1}The saddle point problem}

The problem of quantized transmission across a saddle-point
constriction with magnetic field has been addressed in the past by
B\"uttiker\cite{10}. B\"uttikers paper was motivated by the
theoretical discussion of quantized conduction steps discovered by
van Wees et al\cite{11} and Wharam et al.\cite{12} in split gate
constrictions of a 2-dimensional electron gas. Since constrictions
in these experiments are electrostatically induced with a pair of
split gates, the potential is a smooth function (without hard
walls) and the bottleneck of the constriction therefore forms a
saddle. At this point our view of the role of potential
fluctuations meets the split gate situation referenced above. In
this context we look for a link between the results of B\"uttiker
for the bottleneck situation at high magnetic fields and our semi
empirical back scattering function $P(\Delta\nu)$.

\begin{figure}
\includegraphics[width=7cm]{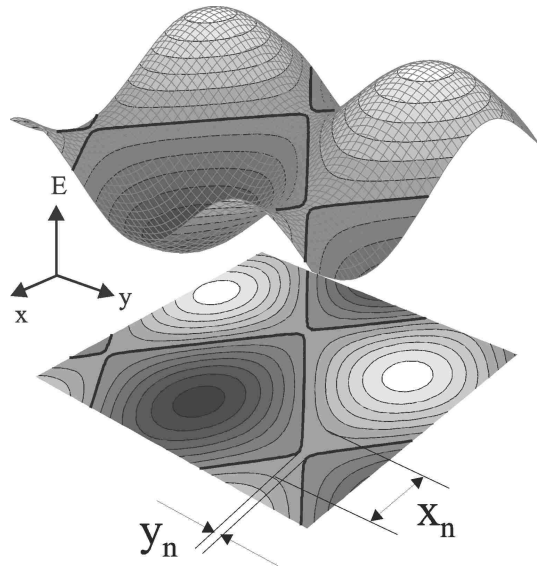}
\caption{\label{Fig1} Example for a two-dimensional potential
modulation which forms saddle points. The minimum distances of the
closed loops (magnetic bound states) are shown for two different
cases: $y_n$ is achieved, if $\varepsilon$ is below the saddle
energy $eV_0$ (bold line) and $x_n$ is achieved, if $\varepsilon$
is above the saddle energy. All possible loops represent
equipotential lines and the selection of the active loops is made
by the position of the Fermi level($\varepsilon = E_F$).}
\end{figure}

Following the paper of B\"uttiker for the high magnetic field
limit, the trajectories of the states are given by equipotential
lines defined by $\varepsilon  = eV(x,y)$, where $\varepsilon$
denotes the energy of the guiding center of the classical
cyclotron orbit, $V$ the potential and $e$ the electron charge. If
$V_0$ denotes the potential of the saddle, the case $\varepsilon <
eV_0$ describes a trajectory, which is repelled by the saddle,
where the energy $\varepsilon$ is finally represented by the Fermi
level. Using the considered saddle point as the center of our
coordinate system, the minimum distance $y_n$ between trajectories
of the same energy, which are classically repelled on opposite
sides to the saddle point, is determined by $\varepsilon = eV(0,
y_n)$ for $\varepsilon < eV_0$ (see Fig. \ref{Fig1}). A similar
situation appears for $\varepsilon > eV_0$, but turned around by
90 degrees and the minimum distance $x_n$ is determined by
$\varepsilon = eV(x_n,0)$. According to B\"uttiker, the
transmission probabilities in terms of $x_n$ and $y_n$ in the high
magnetic field limit for the case of $\varepsilon_n>0$ is

\begin{equation}
T_{mn} = \delta_{mn} \left\{1 + \exp \left[ \pi
\frac{\omega_x}{\omega_y} \left(\frac{x_n}{l_B} \right)^{2}
\right]
 \right\}^{-1}
\label{eq1}
\end{equation}

and for the case $\varepsilon_n<0$ is

\begin{equation}
T_{mn} = \delta_{mn} \left\{1 + \exp \left[ \pi
\frac{\omega_y}{\omega_x} \left(\frac{y_n}{l_B} \right)^{2}
\right]
 \right\}^{-1}
\label{eq2}
\end{equation}

 In Eqn.\ref{eq1} and \ref{eq2} the saddle energy has been
defined as zero reference ($V_0=0$). The underlying problem is
basically a tunneling problem and  $\delta_{nm}$ tells us that
this tunneling process with the tunneling rate $T_{mn}$ happens
mainly between the same states (states of same LL) on opposite
sides of the saddle. The parameters  $\omega_y$ and $\omega_x$
result from expanding the potential near the saddle point by

\begin{equation}
V(x,y)\approx V_0 + 0.5 m \omega_{x}^{2}x^2 - 0.5 m
\omega_{y}^{2}y^2\label{eq3}
\end{equation}

Now we look for a relation between the transmission rates given
above and the function $P(\Delta\nu)$, which represents the ration
between transmission and reflection $R/T$ of edge states like used
in the network model. If we look for $x_n$ while $y_n=0$
($\varepsilon_n = e(V-V_0)>0$), we can easily calculate $x_n$ from
Eqn. \ref{eq3}:

\begin{equation}
x_{n}^{2} = \frac{2\varepsilon_n}{m\omega_x^2} \label{eq4}
\end{equation}

In the case $\varepsilon_n < 0$ we get something analogous for
$y_n^2$:

\begin{equation}
y_{n}^{2} = \frac{2\varepsilon_n}{m\omega_y^2} \label{eq5}
\end{equation}

We introduce an artificial periodic potential modulation, which is
a 2-dimensional Cosine - function and which has the same Taylor
expansion like Eqn.\ref{eq3}:

\begin{equation}
V(x,y) = \tilde{V}  \left[ \cos(\omega_yy) -
\cos(\omega_xx)\right] \label{eq6}
\end{equation}

with $\tilde{V}$ being the potential modulation. Fig. \ref{Fig2}
shows a contour plot of the 2-dimensional Cosine potential
modulation and we get a saddle point whenever the maximum of the
Cosine in one direction meets a minimum of the Cosine in the other
direction. In this way we get a periodic array of saddle-points.
We consider one saddle to be located exactly at the origin of our
coordinate system. If we think about a smoothly varying random
fluctuation potential, the average curvature in both directions
will be the same. In our representative potential in Eqn.
\ref{eq6} we therefore simplify by setting
$\omega_x=\omega_y=\omega$, which leads to $\omega_x / \omega_y =
1$. In order to re-write Eqn. \ref{eq1} and Eqn. \ref{eq2} we do
some further substitutions. For the magnetic length $l_B$ we use:

\begin{figure}
\includegraphics[width=7cm]{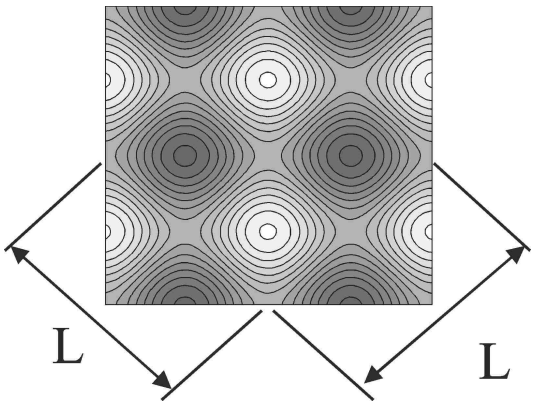}
\caption{\label{Fig2} Contour plot of the artificial 2D Cosine
potential. Dark indicates potential minima and the light colored
areas indicates the potential maxima. $L$ is the period of the
potential modulation and at the same time it is the grid period of
the network.}
\end{figure}

\begin{equation}
l_B^2 = \frac{h}{2\pi e B}  \label{eq7}
\end{equation}

with $h$ the Plank constant, $e$ the electron charge and $B$ the
magnetic field. Using further the period length $L$ of the
periodic arrangement of saddle points

\begin{equation}
L = 2\pi / \omega \label{eq8}
\end{equation}

we can rewrite Eqn. \ref{eq4} and Eqn. \ref{eq5}:

\begin{equation}
x_n^2 = \frac{L^2\varepsilon_n}{2 \pi^2 e\tilde{V}} \label{eq9}
\end{equation}

\begin{equation}
y_n^2 = \frac{L^2\varepsilon_n}{2 \pi^2 e\tilde{V}} \label{eq10}
\end{equation}

Eqn. \ref{eq7} - \ref{eq10} now allow to re-write  Eqn. \ref{eq1}
and Eqn. \ref{eq2}: Since $x_n$ and $y_n$ are described exactly by
the same function, but distinguished only by the case
$\varepsilon_n$ either above or below the saddle energy, we can
rewrite Eqn. \ref{eq1} and Eqn. \ref{eq2} into a single equation
as follows:

\begin{equation}
T_{mn} = \delta_{mn} \left\{1 + \exp \left[
\pm\frac{L^2\varepsilon_n}{e\tilde{V}}\frac{e B}{h} \right]
 \right\}^{-1}
\label{eq11}
\end{equation}

The signes $\pm$ indicate the cases $\varepsilon_n$ above or below
the saddle energy. At this point it should be mentioned, that
already Haug et al \cite{14} have used a similar equation for
addressing the problem of transport across magnetic bound states.
They investigated experimentally the effect of a gate electrode
across the current path of a QHE sample at constant magnetic
field. The major point of that experiment was to study the special
situation of having integer filling outside the gate (pure EC
transport) and sweeping the gate voltage through non-integer
filling below the gate. The authors describe the scattering
through the bulk due to the presence of the gate by partly EC back
scattering at the gate and partly back scattering via tunneling
across magnetic bound states in the gate region. They achieved
good agreement with the experiments already by a simplified
network consisting of just 2 or 3 magnetic bound states.

In order to be able to compare with the formulation of edge
channel back scattering in terms of $P(\Delta\nu)$ , we have to
realize, that the transmission process across the saddle $T_{mn}$
is formally a back scattering process in the EC picture. Therefore
the transmission factor $T_{mn}$ in Eqn. \ref{eq11} corresponds to
a reflection $R$ in the edge channel picture. We transform Eqn.
\ref{eq11} into $P = R/T$ by using $T_{mn}$ as $R$ and
consequently using $(1-T_{mn})$ as $T$. :

\begin{equation}
P = \frac{R}{T}= \frac{T_{mn}}{1-T_{mn}}
 \label{eq12a}
\end{equation}

 Since the above tunneling process happens
between edge states on opposite side of the saddle but the same
Landau level (LL), $\delta_{mn}$ can be omitted for the further
treatment. Further more we have to consider, that the energy
$\varepsilon_n$ is determined by the Fermi level $E_F$. In this
way we end up with $P$ as an exponential function of the Fermi
level $E_F$:

\begin{equation}
P = \exp \left[ \mp\frac{L^2 E_F}{e\tilde{V}}\frac{e B}{h} \right]
 \label{eq12}
\end{equation}

In this equation $E_F$ is the Fermi level relative to the saddle
energy. The ration $L^2 / \tilde{V}$ can be understood as a
measure of the "smoothness" of the potential modulation and $e B /
h$ is the well known number of states in a single LL per unit
area. Since $L$ is the distance between equivalent saddle points,
$L^2$ is the area of the 2-dimensional basis cell of the network
grid. It is interesting to note, that the pre-factor of $E_F$,
which is $(L^2 / e\tilde{V}) (e B / h)$, has the meaning of an
average density of states (DOS) in the basis cell of the network,
provided that we associate the LL broadening with $e\tilde{V}$. In
order to consider this aspect in a little more detail, we assume a
Gaussian shaped density of states for a single LL \cite{16} and
calculate the DOS in the center of the LL as follows:

\begin{equation}
DOS(\varepsilon_n = 0) = \frac{e B}{h} \frac{1}{b \sqrt{\pi}}
 \label{eq13}
\end{equation}

where $b$ is the energy broadening of the Gaussian peak. On this
background we can calculate the change of the carrier density
$\Delta n$ in a single LL due to the change of $E_F$ by  $\Delta
E_F$ as follows:

\begin{equation}
\Delta n = \frac{e B}{h} \frac{1}{b \sqrt{\pi}}\Delta E_F
 \label{eq14}
\end{equation}

This is valid for a Fermi level close to the center of the LL,
where the DOS is approximately constant. However, we can also
represent $\Delta n$ in terms of the filling factor by $\Delta \nu
= \Delta n / n_{LL} $, where $n_{LL} = e B / h$. This results in:

\begin{equation}
\Delta \nu = \frac{\Delta E_F}{b \sqrt{\pi}}
 \label{eq15}
\end{equation}

Using again the saddle energy as the zero reference, $\Delta E_F$
becomes $E_F$ and $\Delta \nu$ becomes the difference of the
filling factor relative to half filling. Now we can substitute
$E_F$ by $\Delta \nu$ and rewrite Eqn. \ref{eq12}:

\begin{equation}
P = \exp \left [ - \Delta \nu \frac{b \sqrt{\pi}}{e\tilde{V}} L^2
\frac{e B }{h} \right ]
 \label{eq16}
\end{equation}

Since $\tilde{V}$ is the amplitude of our artificial potential
modulation and $b$ is the with of the representative DOS - peak,
these two parameters correspond to each other. Therefore we can
assume, that the pre-factors in the exponent will be a factor of
the order of unity, just depending on details of the real
potential fluctuations. On this basis we simplify

\begin{equation}
\frac{b \sqrt{\pi}}{e\tilde{V}} \approx 1
 \label{eq17}
\end{equation}

and get the following exponential function:

\begin{equation}
P = \exp \left [ - \Delta \nu L^2 \frac{e B }{h} \right ]
 \label{eq18}
\end{equation}

It is easily seen, that the semi-empirical exponential function of
Ref. \cite{3} is completely recovered, except the fact, that we
get a more complex pre-factor for the filling factor $\Delta \nu$.
The explicit magnetic field dependence of the pre-factor is a new
feature as compared to the earlier semi-empirical function.
However, this has no impact on the major features of the function
$P(\Delta \nu)$, just causing a magnetic field dependence of the
width of the plateau transition regimes. If we take a closer look
to the exponent, we find $e B/h$, which is the number of states
per unit area in a LL and $L^2$, which is the area of a grid
period of the network. As a consequence we get a surprisingly
universal result: The exponent of the function $P(\Delta \nu)$
represents the number of carriers being added ($\Delta N > 0$) or
removed ($\Delta N < 0$) from the half filled basis cell of the
network grid. This happens if the Fermi level moves away from the
center position of the LL.

\begin{subequations}
\label{eq19}
\begin{equation}
 P = \exp \left [ - \Delta N \right ]
 \end{equation}
\begin{equation}
\Delta N = \Delta \nu L^2 \frac{e B }{h}
\end{equation}
\end{subequations}

This is usually achieved by changing the magnetic field at
constant carrier density. However, this change normally happens
continuously, meaning that the number of states in a single LL in
one grid period is changed just in fractions of unity. This is
possible, because in the regime close to a half filled LL there is
a strong coupling between the individual grid periods so that the
(single) electronic states are well de-localized and can be
smeared over several grid periods. A change of the number of
electronic states by $\Delta N = \pm 1$ corresponds to a complete
plateau to plateau transition. By considering the grid period
length $L$ as a measure of the smoothness of the potential
fluctuations, we get a nice possibility for an interpretation of
our result: A large $L$ corresponds to a rather clean sample (high
mobility) with very smooth potential fluctuations and a short $L$
corresponds to a sample with strong potential fluctuations and
hence, strong disorder (low mobility). As a consequence, a large
$L$ indicates a large area of the basis cell of the network and
therefore only a small change of the filling factor $\Delta \nu$
and thus, only a small change of the magnetic field is needed in
order to change the number of states by one. This means that we
get sharp plateau to plateau transitions on the magnetic field
axis. In contrast, a small $L$, as expected from samples with
strong disorder, leads to a small area of the basis cell, which
then requires a large change of the filling factor $\Delta \nu$
for changing the number of states by the order of one. Therefore
we get broad plateau to plateau transitions in this case, exactly
what we expect for samples with strong disorder.

\begin{figure}
\includegraphics{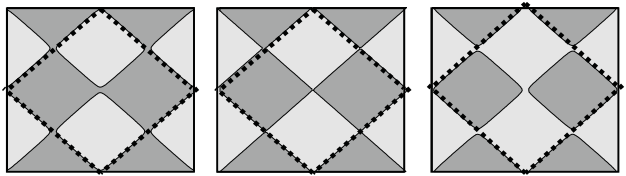}
\caption{\label{Fig3} Spatial distribution of filled (dark
colored) and empty (light colored) states for left: $E_F >
E_{LL}$,middle: $E_F = E_{LL}$ and right: $E_F < E_{LL}$. The
dotted line indicates the basis cell of the network.}
\end{figure}

On the basis of geometric arguments and using a checkerboard as
representative for the network (Fig. \ref{Fig3}) Eqn. \ref{eq19}
allows also another simple interpretation: Using directly the
trajectories of the channels at different Fermi levels we get 3
cases as shown in Fig. \ref{Fig3}: (i) if the Fermi level is above
the saddle energy (center of LL), the filled states dominate the
transport properties, because they build an interconnected
network. (ii) if the Fermi level coincides with the saddle energy,
filled and empty states take up exactly half of the space, which
also means exactly a half filled LL. (iii) if $E_F$ is below the
saddle energy, the empty states (insulating phase) build an
interconnected network, leaving isolated droplets of filled states
and leading to a high resistive regime. It should be mentioned,
that already Mashida et al. used such a picture for the
interpretation of their experimental results\cite{Komiyama2,
Komiyama3}.

\section{\label{sec:level1}Discussion}

\begin{figure}
\includegraphics{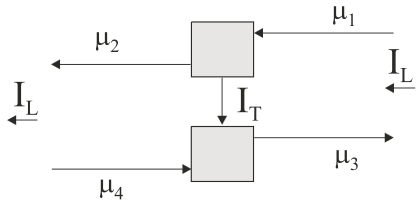}
\caption{\label{Fig4} Schematic representation of the effect of
back scattering in the EC-picture.}
\end{figure}

By using a representative 2-dimensional Cosine potential
modulation, a result by B\"uttiker \cite{10} for the saddle point
problem at high magnetic fields was transformed into the back
scattering function $P(\Delta \nu)$, which is used for the nodes
of the network according to Ref. \cite{network}. One main
extension as compared to Ref. \cite{10} is the treatment of the
tunneling current within the framework of the Landauer-B\"uttiker
formalism. This means that a transverse current between a channel
pair (tunneling current serves as a back scattering current
between channels), leads to a longitudinal potential drop in each
of the channels. This can be considered as the essence of the
EC-picture within the LB-formulation and there is no classical way
to account for this behavior. In order to point out this fact
somewhat more clearly, the situation is sketched in Fig.
\ref{Fig4}.

If there is a longitudinal current $I_L$, it is carried
dissipation-less by a channel pair and is represented on the right
side by $I_L = (\mu_1 - \mu_3)e^2/h$ and on the left side by $I_L
=(\mu_2 - \mu_4)e^2/h$, while $(\mu_1 - \mu_3) = (\mu_2 - \mu_4)$
for current conservation reasons. A transverse current $I_T$,
which may be enabled by some conduction or tunneling between
opposite channels causes a longitudinal voltage drop. This is
because $I_T$ has to be represented also within the LB formalism
according to $I_T = (\mu_1 - \mu_2)e^2/h$ on the upper edge and
$I_T = (\mu_3 - \mu_4)e^2/h$ at the lower edge, while $(\mu_1 -
\mu_2) = (\mu_3 - \mu_4) = \Delta \mu_L $, again because of
current conservation reasons. It is important to note, that any
current ($I_L$ or $I_T$) within the EC-picture is defined only by
a pair of directed channels, which must not be confused with a
difference of opposite directed classical currents. As a
consequence, it is not possible to assign a current to any single
directed channel and the occurrence of a perpendicular current
$I_T$ does not change the longitudinal current $I_L$. This is a
fact, which must not be misinterpreted as a violation of
Kirchhoff's law of current conservation. Using finally the ration
$I_T / I_L$ as an equivalent for $R / T$ of the LB formalism, it
is easily shown that the longitudinal voltage drop $\Delta \mu_L$
is coupled to the Hall voltage like $\Delta \mu_L = (\mu_1 -
\mu_2) = (\mu_1 - \mu_3) R / T$, with $R/T$ identified as $P$,
which is the function derived in this paper.

\begin{figure}
\includegraphics{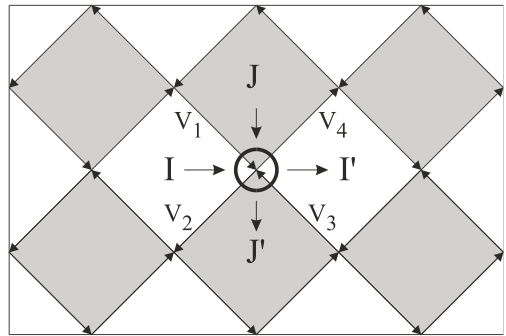}
\caption{\label{Fig5} Checkerboard model for bulk current
transport in the QHE regime after Ruzin et al \cite{9}. The dark
and light areas represent different quantum Hall liquids with
different Hall conductance. Currents I and J are assumed to flow
exclusively in the corresponding phase and the total behavior of
the conductor is determined by the mixture of both phases. }
\end{figure}

The above summarized view of the EC-picture has been used already
in Ref. \cite{17} in order to explain an anomalous behavior of the
magneto transport of high mobility quasi 3-dimensional PbTe wide
quantum wells. Also in Ref. \cite{3} this interpretation of the
EC-picture has been successfully used which finally led to the
application within the discussed network model. From this point of
view the network model can be considered to be based purely on the
EC-picture and the LB-formalism. It is therefore quite surprising,
that also the ohmic regime is quite well represented. However,
this puzzle can be resolved by reconsidering Fig. \ref{Fig3},
which appears like a checker board representation of the network.
From this point of view it meets the approach of Ruzin et al
\cite{9}. Ruzin et al used such a checker board model for bulk
transport in the QHE regime in order to obtain a universal
relation between longitudinal and transverse conductances in the
QHE and Fig,5 summarizes this concept: Ruzin et al considered two
competing quantum Hall liquids in the presence of a long-range
random potential, with the magnetic field determining the area
fractions close to $1/2$. Fig. \ref{Fig5} represents this
situation for an artificial periodic long-range potential. The
"white" regions represent the phase with quantized Hall
conductance  $\sigma_1$, and the "dark" regions those of Hall
conductance $\sigma_2$. If the situation is such that the "white"
phase percolates freely throughout the sample, the system will be
on the plateau with  $\sigma_{xy} = \sigma_1$. Similarly if the
"dark" regions percolate freely, the system will be on the plateau
with $\sigma_{xy} = \sigma_2$. Near the percolation threshold
transport is controlled by quantum tunneling between clusters of
the same phase at the saddle points. Ruzin et al related the net
currents in the four quadrants to the four electric potentials at
the edges $V_1$ through $V_4$ as given by:

\begin{subequations}
\label{eq20}
\begin{equation}
 I = \sigma_1 (V_2 - V_1)
 \end{equation}
\begin{equation}
 J = \sigma_2 (V_3 - V_2)
 \end{equation}
 \begin{equation}
 I' = \sigma_1 (V_3 - V_4)
 \end{equation}
 \begin{equation}
 J' = \sigma_2 (V_4 - V_1)
 \end{equation}
\end{subequations}

\begin{figure}
\includegraphics{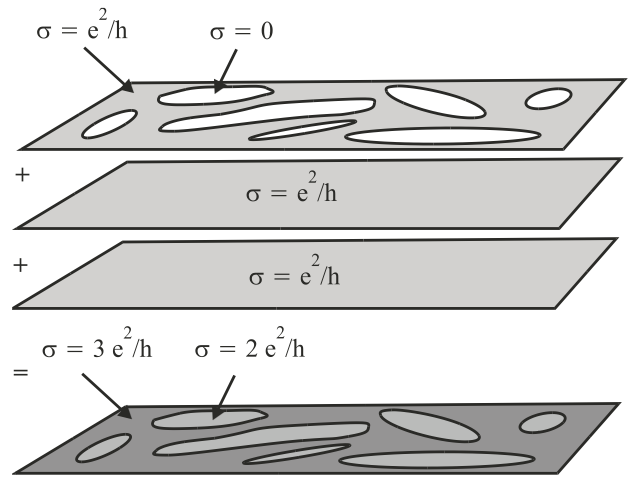}
\caption{\label{Fig6} Scheme for combining individual LLs, where
each of them may have a mixture of quantum Hall liquid and
insulating phase. In this case it is assumed that just the top LL
is partly occupied and all lower ones are completely filled. The
result is a combined system of mixed phases of different quantum
Hall liquids, which means different Hall conductivities.}
\end{figure}

Postulating current conservation within the individual phases ($I
= I'$ and $J = J'$), Ruzin et al get automatically $(V_2-V_1) =
(V_3 - V_4)$, which is identical to the results of our treatment
as illustrated in Fig. \ref{Fig4}.  Further on, Ruzin et al
obtained a universal semi-circle-relation between longitudinal and
transverse conductance. At this point it should be mentioned, that
such a semicircle relation has also been obtained within our
preceding work \cite{3,4}, which once more demonstrates the
agreement of our results with the results of Ruzin et al. However,
our model achieves additional results such as the filling factor
dependence and transport parameters in terms of $R_{xx}$ and
$R_{xy}$ as well as the distribution of the applied voltage across
the sample area. There is also a qualitative difference between
Ruzin's and our picture: In our picture the space is divided
between occupied and non-occupied states, while Ruzin et al are
talking about 2 competing quantum Hall liquids (QHL) of different
Hall conductivity. We can overcome this disagreement by
recognizing, that in our model each involved LL is represented by
a separate network with its own mixture of empty and filled
states, which we associate with the insulator and a QHL phase of
Hall conductance $\sigma = e^2/h$ respectively. In the case of the
IQHE there is always just one partly filled (top) LL and all lower
ones are completely filled. For each LL the QHL phase exhibits a
quantized Hall conductance of $\sigma = e^2/h$, while the Hall
conductance for the insulator phase $\sigma = 0$. For the complete
system in the bulk current representation we have to sum up all
contributions of all LLs locally, which leads to a situation like
sketched schematically in Fig. \ref{Fig6}. Putting forward a bulk
current representation, one has to add all Hall conductivities of
all involved LLs locally and the resulting picture becomes again
similar to that one of Ruzin et al. At this point we want to make
clear, that at this stage our network model does not include
non-linear effects, which means that it considers the electron
system to be sufficiently close to thermal equilibrium. However,
for thermal equilibrium some similarity between bulk and edge
current picture might not be that surprising. A recent theoretical
treatment of non-linear transport through mesoscopic systems by
Sanchez et al \cite{Buttiker_asymm} shows a possible magnetic
field asymmetry, which is not within the scope of our network
model in the present state of development. Nevertheless it is
possible to describe selective current injection into edge stripes
also within our network model, like e.g. achieved by selective EC
reflection due to gate electrodes\cite{network}, while we cannot
see how this could be achieved within the bulk current picture.
Another important aspect to be mentioned at this point is that in
our network the phase coherence is not maintained over more than
one single grid period, but still exact quantization of the QHE is
obtained. This is in agreement with theoretical results of F.
Gagel et al \cite{Gagel} who found a surprising stability of the
QH plateaus against dissipation and phase destroying events.

\section{\label{sec:level1}Simulation results}

\begin{figure}
\includegraphics{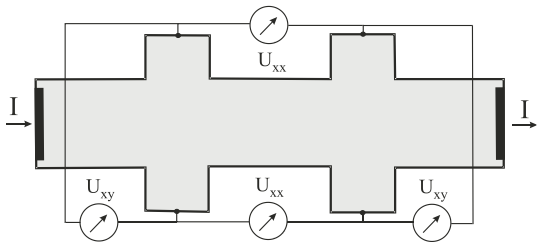}
\caption{\label{Fig7} Sample layout as used for the numerical
Simulations.}
\end{figure}

\begin{figure}
\includegraphics[width=8cm]{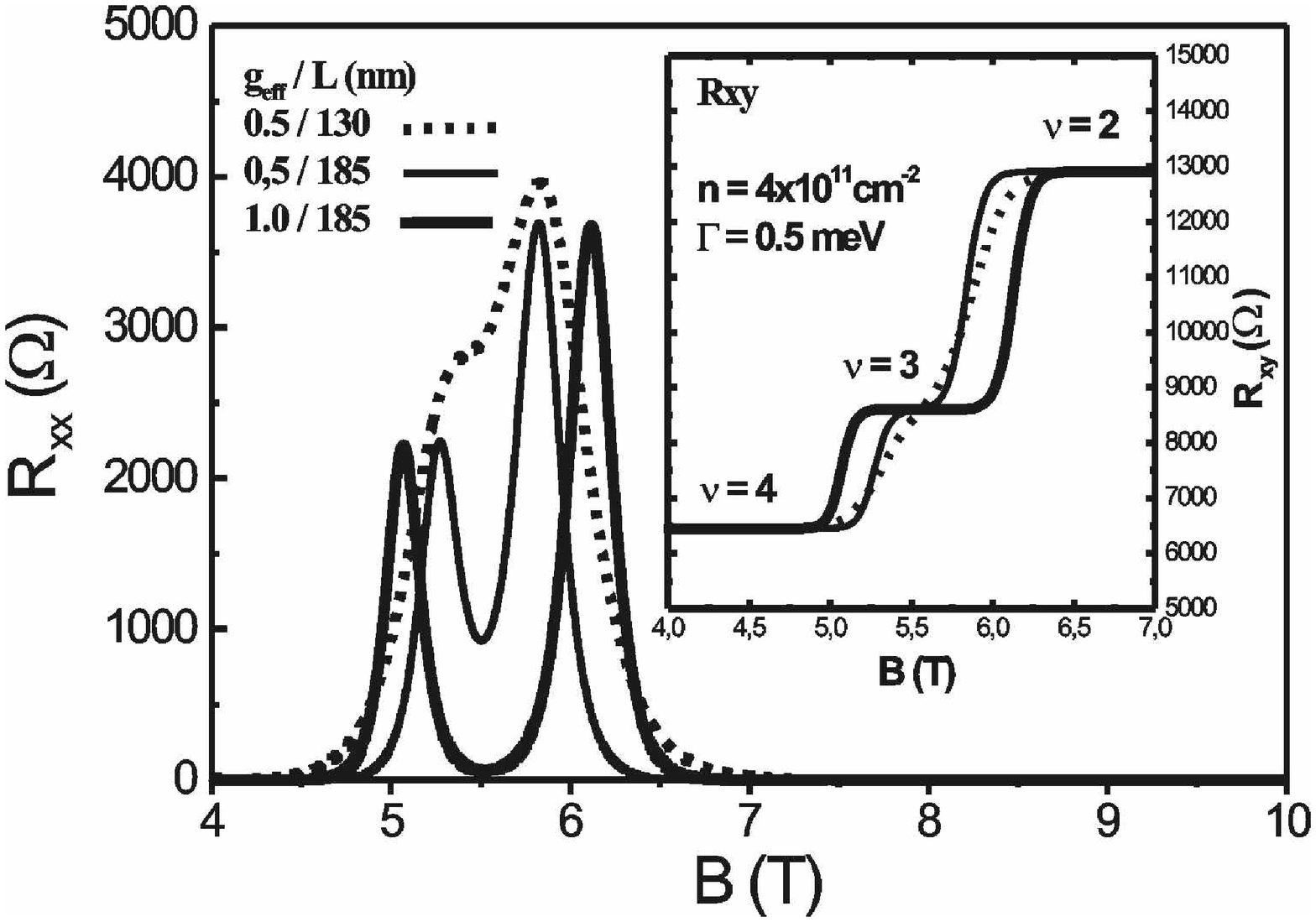}
\caption{\label{Fig8} Results for magneto transport data in the
high field regime calculated with Eqn. \ref{eq18} as the coupling
function for the nodes of the network. The calculation is done for
2 different effective g-factors ($g_{eff} = 0,5$ and $g_{eff} =
1.0$) and two different values for $L = 130nm$ and $L = 185nm$. }
\end{figure}

\begin{figure}
\includegraphics[width=8cm]{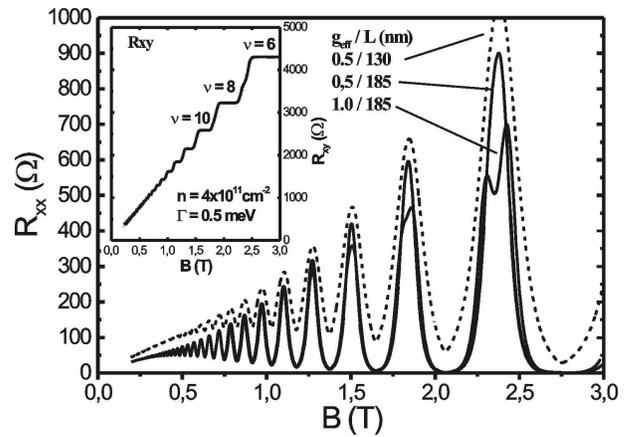}
\caption{\label{Fig9} Results for magneto transport data in the
low field regime calculated with  Eqn. \ref{eq18} as the coupling
function for the nodes of the network. The calculation is done for
2 different effective g-factors ($g_{eff} = 0,5$ and $g_{eff} =
1.0$) and two different values for $L = 130nm$ and $L = 185nm$. }
\end{figure}

If we use Eqn. \ref{eq18} as a replacement of the original
function of Ref. \cite{network}, we get results for the magneto
transport simulation as shown in Fig. \ref{Fig8} and Fig.
\ref{Fig9}. The layout of the sample is shown in Fig. \ref{Fig7}.
Considering the individual plateau transitions at high magnetic
fields, there is no qualitative difference to the results of the
previous version. The major difference and improvement concerns
the low magnetic field range. At very low magnetic fields the
zeros in $R_{xx}$ disappear and the Hall effect starts as a
straight line, and turns gradually over to a plateau behavior at
higher fields. The reason for this is the explicit magnetic field
dependence of the exponent in Eqn. \ref{eq18}, which leads to wide
plateau transitions at low magnetic fields and sharp plateau
transitions at high magnetic fields. This results in an overlap of
the ohmic (dissipative) regimes of several LLs at low fields. In
order to get the individual filling factors in case of LL overlap,
the individual spin split LLs have been calculated using the
standard parameters for GaAs and a LL broadening according to
$\Gamma = \Gamma_0 \sqrt{B}$ \cite{16} has been used by setting
$\Gamma_0 = 0.5meV$.

The actual calculation consists of two main parts: (I) For
calculating the occupation numbers standard procedures are used
and the Fermi level in the bulk region is calculated by filling up
the density of states (DOS) with the constant bulk carrier
density. The DOS is composed by the superposition of the magnetic
field dependent Gaussian shaped DOS of spin split LLs. The
Fermilevel for regions of non-zero electrostatic bare potential
(edges and gate regions) is forced to match the obtained (magnetic
field dependent) Fermi level in the bulk. This allows to get a
self-consistent electrostatic potential and a rearrangement of the
carrier density at the edges, which results in an electrostatic
edge potential according to Chlovskii et al \cite{Chklovskii}.
(II)The such obtained lateral carrier density profile is
transferred to the nodes of the network and in another
self-consistent iteration procedure the lateral distribution of
the bias voltage, which is introduced via the current contacts, is
calculated. From this the potential difference for any designated
pair of voltage probes is obtained as a function of the magnetic
field. A metallic contact is realized by interconnecting all
channels of all involved LLs at the position of the designated
contact point\cite{network}. Extended metallic contacts are
realized by creating arrays of such contact points. The current at
the current contacts is calculated only after arriving at the
self-consistent solution in the network, which allows finally to
calculate the various resistances. This means, that in principle a
constant supply voltage is used in the network model instead of a
constant supply current like in most experiments. However, for
calculating the resistances for a standard QHE setup with a single
current source this makes no difference. For achieving a constant
current mode, the potentials at the current contacts are
additionally varied in a proper way during the iteration procedure
in order to get the required preset current. In this way a
one-to-one simulation of experiments using constant supply
currents and measuring voltages at several potential probes is
possible.

Just for demonstration, the influence of spin splitting by using
different effective g-factors such as $g^{*} = 0.5$ and $g^{*} =
1.0$ and different values for L, such as L = 135nm and L = 185nm,
is simulated. The results are shown in Fig. \ref{Fig8} for the
high magnetic field range and in Fig. \ref{Fig9} for low magnetic
fields. For $L = 135nm$ and $g^{*} = 0.5$ spin splitting is hardly
to observe even at high fields around 5 Tesla, while for the case
$L = 185nm$ a well resolved spin splitting can be observed even
for a small g-factor of $g^{*} = 0.5$. At magnetic fields around
2.5 Tesla spin splitting is only weakly pronounced even for $g^{*}
=1$ and $L = 185nm$. In summary one can say, that this overall
behavior looks quite realistic and a more systematic study on the
possibility of a more accurate determination of an enhanced
effective g-factor from transport data is planned separately.

\section{\label{sec:level1}Summary}

Starting with a treatment of the saddle point problem in the high
magnetic field limit by B\"uttiker we have been able to show, that
it is possible to transform B\"uttikers results into a filling
factor dependent back scattering function $P(\Delta\nu)$. This
function is used as the coupling function for the nodes of a
recently developed new network model for the QHE regime of 2D
systems. As a result, the numerical simulation based on this
network delivers transport data, in which  $R_{xy}$ starts with a
classical straight line while $R_{xx}$ shows no zeros. With
increasing magnetic field the IQHE sets-in with Shubnikov-de Haas
oscillations in $R_{xx}$, followed by zeros in $R_{xx}$ and
plateaus in $R_{xy}$, like seen in experimental data. For
equilibrium conditions we have found an equivalence between the
bulk current picture in terms of mixed phases like used by Ruzin
et al and our network representation on the basis of the
Landauer-B\"uttiker formalism.

\section{\label{sec:level1}Acknowledgements}

The author thanks M. B\"uttiker for stimulating discussions and
comments.

\end{document}